\documentclass[11pt,preprint,flushrt]{aastex}
\usepackage{amsmath}
\usepackage{amssymb}
\usepackage{latexsym}
\usepackage{cancel}
\usepackage{verbatim}
\usepackage{indentfirst}
\usepackage{multirow}
\usepackage{color}

\usepackage{natbib}

\usepackage{graphicx}

\begin{document}

\def\dr{\mbox{$\Delta \overline{R}$}}
\def\dv{\mbox{$\Delta V$}}

\title{Atmospheric dispersion effects in weak lensing measurements}
\author{Andr\'es Alejandro Plazas \& Gary Bernstein}
\affil{Department of Physics and Astronomy, University of Pennsylvania, Philadelphia, PA, 19104}

\begin{abstract}
The wavelength dependence of atmospheric refraction causes elongation of finite-bandwidth images along the elevation vector, which produces spurious signals in weak gravitational lensing shear measurements unless this atmospheric dispersion is calibrated and removed to high precision.  Because astrometric solutions and point spread function (PSF) characteristics are typically calibrated from stellar images, differences between the reference stars' spectra and the galaxies' spectra will leave residual errors in both the astrometric positions (\dr) and in the second moment (width) of the wavelength-averaged PSF (\dv) for galaxies.  We estimate the level of \dv\ that will induce spurious weak lensing signals in PSF-corrected galaxy shapes that exceed the statistical errors of the {\em Dark Energy Survey (DES)} and the {\em Large Synoptic Survey Telescope (LSST)} cosmic-shear experiments.  We also estimate the \dr\  signals that will produce unacceptable spurious distortions after stacking of exposures taken at different airmasses and hour angles.
Using standard galaxy and stellar spectral templates we calculate the resultant errors in the $griz$ bands, and find that atmospheric dispersion shear systematics, left uncorrected, are up to 6 and 2 times larger in $g$ and $r$ bands, respectively, than the thresholds at which they become significant contributors to the {\em DES} error budget, but can be safely ignored in $i$ and $z$ bands. For the stricter {\em LSST} requirements, the factors are about 30, 10, and 3 in $g, r$, and $i$ bands, respectively. These shear systematic errors scale with observed zenith angle $z$ as $\langle \tan^2 z \rangle$, for which we take a nominal value of unity---simulations of {\em DES} and {\em LSST} suggest 0.6--1.0. 
We find that a simple correction linear in galaxy color is accurate enough to reduce dispersion shear systematics to insignificant levels in the $r$ band for {\em DES} and $i$ band for {\em LSST}, but still as much $5\times$ above the threshold of significance for {\em LSST} $r$-band observations.  More complex approaches to correction of the atmospheric dispersion signal will likely be able to reduce the systematic cosmic-shear errors below  statistical errors for LSST $r$ band. But $g$-band dispersion effects remain large enough that it seems likely that induced systematics will dominate the statistical errors of both surveys, and cosmic-shear measurements should rely on the redder bands.

\end{abstract}

\section{Introduction}
Weak gravitational lensing (WL) can be used for high-precision measurements of the expansion history of the Universe and the evolution of gravitational potentials within it \citep{mellier99,refregier03,hoek_jain08}. The WL effect by large-scale structure is best detected as a coherent elongation of background galaxy images, ``cosmic shear,'' which induces an RMS shift in axis ratios of only $\approx 1\%$ red or less for cosmologically distant source galaxies.  
Several ground-based surveys of $>1000$~deg$^2$ are commencing this year, including the $5000$~deg$^2$ {\em Dark Energy Survey} (DES)\footnote{http://www.darkenergysurvey.org/}, with the goal of measuring the amplitude and redshift dependence of cosmic shear to high precision.  A yet larger and deeper survey project for late in this decade, the {\em Large Synoptic Survey Telescope} (LSST)\footnote{http://www.lsst.org/lsst/}, is under intensive development.  

Ground-based visible observations are subject to atmospheric refraction, shifting each photon toward the zenith by an angle $R(\lambda, z_a)$ that depends on wavelength and on apparent zenith angle $z_a$.  For finite-bandwidth observations, there are two measurable consequences: first, a centroid shift
\begin{equation}
\overline{R} = \frac{\int d\lambda R(\lambda,z_a) F(\lambda) S(\lambda)}{\int d\lambda F(\lambda) S(\lambda)}
\label{r}
\end{equation}
that depends on $z_a$, the instrumental response function $F(\lambda)$, and the source spectrum $S(\lambda)$.  Additionally the dispersion stretches the image along the zenith, which can be quantified by the photon-weighted second central moment of the dispersion kernel:
\begin{equation}
V \equiv \frac{\int d\lambda \left[R(\lambda,z_a)-\overline{R}\right]^2 F(\lambda) S(\lambda)}{\int d\lambda F(\lambda) S(\lambda)}
\label{v}
\end{equation}
Like any other instrumental distortion or convolution of the galaxy images, this coherent vertical elongation of the galaxy images will be mistaken for WL shear if not removed.  The standard strategy in WL observing is to use stellar images to estimate the instrumental point spread function (PSF), interpolating the stars' PSFs to the location of each galaxy, and then effect some limited form of deconvolution to extract the ``pre-seeing'' shape of the galaxy free of instrumental effects.  In its simplest form, this instrumental correction amounts to subtracting the intensity-weighted second moments of the PSF from those observed for the galaxy to estimate the pre-seeing moments.  If the atmospheric dispersion of the galaxies and stars are the same, then this stellar PSF correction will automatically correct the galaxy images for the dispersion $V$.\footnote{To be precise, the dispersion acts as a convolution, like the PSF, only if the galaxy's shape is constant across the band, {\it i.e.\/} the galaxy has homogeneous color.  We will ignore in this work the complications of combining atmospheric dispersion with color gradients, as they produce shear errors at lower order.}  If, however, a galaxy's $V$ differs by an amount $\dv$ from the mean stellar $V$ used to estimate the PSF, then the inferred galaxy shape will be incorrect and propagate to cosmic shear errors.   Our goal is to calculate the size of this effect, see if it will contribute a significant error to the cosmological results of DES or LSST, and evaluate some rudimentary techniques for eliminating the problem.

Since cosmic-shear measurements are not affected by small shifts in the {\em positions} of galaxy images, we are less concerned with the centroid shift $\overline{R}$ of the galaxy image than with the broadening $V$.  Suppose, however, that we intend to combine information from exposures at multiple hour angles and/or filters, for instance by stacking the images before measuring shapes.  The registration of images will normally be done by forcing overlap of stellar images.  Spectral differences between the galaxy and the mean calibrating star will induce a difference \dr\ in their mean refractions which depends on filter and hour angle.  The stacked image will register the stars properly but not the galaxy, broadening the stacked galaxy image and hence producing a coherent shape error which again is confused with cosmic shear.  Since both DES and LSST plan to combine many exposures to produce a single galaxy shape, we estimate \dr\ and determine whether it causes cosmic-shear errors comparable to the statistical errors of these surveys. This problem can be avoided by leaving the galaxy centroid free to differ among exposures in a shape analysis---but this typically requires that the galaxy have high signal-to-noise ratio in a single exposure, which precludes use of most of the galaxies detectable with the full combined survey data.

The effect of atmospheric dispersion on weak lensing shear measurements was crudely estimated by \citet{ksb}, who determined that it should be unimportant to the measurement of galaxy cluster masses.  In the intervening 15 years, however, WL observers have become far more ambitious, aiming to measure the much smaller cosmic-shear signal to high precision, so it is necessary to re-evaluate these limits, especially since neither the DES nor LSST optics incorporate atmospheric dispersion correctors.  \citet{TysonJee} investigate the impact of atmospheric dispersion on the LSST observations, but only with the criterion that the dispersion $V$ be a small contributor to the overall size of the PSF---they did not investigate the much stricter criterion that $V$ be correctable to a small fraction of the expected cosmic-shear signal.

In the next section we derive requirements on \dr\ and \dv\ such that they will not bias cosmic-shear measurements at the level of the statistical errors of the DES and LSST surveys.   Then \S\ref{calculations} describes our methods for estimating these quantities, and \S\ref{results} gives the resultant values in the $griz$ filters for a range of galaxy types and redshifts.  Having demonstrated that atmospheric dispersion is indeed a significant issue in all but the $z$ band, we ask in \S\ref{corrections} whether simple color-based corrections to \dr\ and \dv\ can recover the required accuracy.  We conclude with an outlook in \S\ref{conclusion}.

\section{Tolerable levels of atmospheric dispersion}
\label{effects}
\subsection{Effect of dispersion on measured shear}
We use a simplified model of WL shear measurement to make rough estimates of the tolerable levels of \dr\ and \dv\ for cosmic-shear experiments.  Galaxies are assigned shapes via their intensity-weighted second central moments
\begin{eqnarray}
I_{\mu\nu} & \equiv & \frac{1}{f}\int dx\,dy\, I(x,y) (\mu - \bar \mu)(\nu - \bar \nu), \\
\bar \mu & \equiv & \frac{1}{f}\int dx\,dy\, I(x,y) \mu, \\
f & \equiv & \int dx\,dy\, I(x,y)
\end{eqnarray}
where $I(x,y)$ is the surface brightness distribution of the source.  In practice these integrals have divergent noise so must be bounded by some window function, which complicates the measurement but has little impact on our estimates of the effect of the (small) atmospheric dispersion on the measured shapes.  The ellipticities are assigned as 
\begin{eqnarray}
e_1 & \equiv & \frac{I_{xx} - I_{yy}}{I_{xx}+I_{yy}} \\
e_2 & \equiv & \frac{2I_{xy}}{I_{xx}+I_{yy}}.
\end{eqnarray}
and we can also assign the galaxy a second-moment-based radius, usually quite similar to its half-light radius,
\begin{equation}
r^2  \equiv  \left( I_{xx}+I_{yy} \right).
\end{equation}
The two components $(\gamma_1,\gamma_2)$ of the applied gravitational lensing shear are estimated as $2\gamma_i \approx \langle e_i \rangle$, where the average is taken over the ensemble of galaxies in a selected region of sky position and source redshift.  

These formulae require access to the galaxy's intensity $I^g(x,y)$ {\em before} any instrumental distortions, but we only have access to the observed $I^o$ {\em after} convolution with the PSF (including atmospheric dispersion components) $I^\star$.  For present purposes it is a good approximation that: 
\begin{enumerate}
\item The second central moments of the observed image are the sum of those from the galaxy and the PSF, which means that the galaxy  moments can be estimated as $I^g_{\mu\nu} = I^o_{\mu\nu} - I^\star_{\mu\nu}.$
{\color{black} This holds exactly for shape measurements with unweighted second moments, which are not practical.  However the additivity of second moments under convolution also holds for Gaussian galaxies and PSFs, hence we can expect this relation to be a good approximation for shape measurements based on Gaussian-weighted moments, such as \citet{ksb}.}
\item The effect of atmospheric dispersion is to boost slightly the second moment of the PSF in the zenith direction (call this the $x$ direction), $I^\star_{xx}\rightarrow I^\star_{xx}+V$.  {\color{black} This relation is exact for unweighted second moments, but is also a very good approximation for Gaussian PSFs in the limit when the atmospheric dispersion is much smaller than the PSF itself---which is certainly true here.}
The same boost is given to $I^o_{xx}$, and the subtraction will cancel $V$ in the estimate of intrinsic moments if our PSF is correctly calculated for the galaxy.
\end{enumerate}
Under these conditions we see that a small error $\dv \ll r^2$ in the estimate of the $V$ for a galaxy's PSF will result in an error in the zenith second moment, $I^g_{xx}\rightarrow I^g_{xx}+\dv$, propagating to shape errors
\begin{eqnarray}
e_1 & \rightarrow & e_1\left(1-\frac {\dv}{r^2}\right) + \frac {\dv}{r^2} \\
e_2 & \rightarrow & e_2\left(1-\frac {\dv}{r^2}\right). 
\end{eqnarray}
The shear estimate on a patch of sky is one-half the average galaxy ellipticity.  In the shear-measurement literature it is typical to distinguish a {\em multiplicative} shear error $m$ and an {\em additive} shear error $c$ as $\gamma_i^{\rm est} = (1+m_i) \gamma_i^{\rm true} + c_i$ \citep{STEP}.  Comparing to the above, we infer that the atmospheric dispersion errors propagate to a multiplicative shear error
\begin{equation}
\label{inducedm}
m = \left\langle \frac{\dv}{r^2} \right\rangle,
\end{equation}
where the average is over the galaxies in a chosen bin of sky area and source redshift.  The dispersion error also produces an additive shear error in the zenith direction of very similar amplitude, $c_1\approx m/2$.

\subsection{Shear measurement error budgets}
The multiplicative error $m$ in shear measurement propagates directly to a fractional error $2m$ in the cosmic-shear power spectrum.  Particularly pernicious are multiplicative shear errors that vary as a function of source redshift, since most of the power of cosmic-shear experiments comes from tomographic analyses of detailed comparison between shear strength on different bins of source redshift.  \citet{HTBJ} (hereafter HTBJ) calculate the degradation of cosmological constraints for the DES and LSST cosmic-shear surveys due to multiplicative shear calibration errors with unknown redshift dependence.    
{\color{black} From their Figure~4 we can infer that a systematic error in $m$ of 0.008 (0.003) for DES (LSST) will raise the experiment's uncertainty in the dark-energy equation-of-state parameter $w$ by a factor of $\sqrt 2$ above the purely statistical errors.  [It takes $\approx5\times$ larger $m$ values to double the cosmological errors above the statistical uncertainties.]}
\citet{AmaraRefregier08} carry out a similar analysis, concluding that $|m|<0.001$ is required for a space-based cosmic-shear survey somewhat more ambitious than LSST, consistent with HTBJ at factor-of-2 level.
Since atmospheric dispersion is just one of several potential sources of shear calibration error, an experimental error budget would likely need to hold the contribution from atmospheric dispersion to $\le \frac{1}{2}$ of the total value in HTBJ.  We can hence set a requirement that
\begin{equation}
\label{dvreq}
\left|  \left\langle \frac{\dv}{r^2} \right\rangle \right| < \left\{
\begin{array}{cc}
0.004 & {\rm (DES)} \\
0.0015 & {\rm (LSST)}
\end{array} \right.
\end{equation}

Placing limits on additive shear errors is more complex because the limits on $c$ depend strongly on how the additive signals correlate across redshift bins.  We can make a crude estimate, however, as follows: first, we note that the RMS value of cosmic shear is $\approx0.02$, so the integrated shear power spectrum is $\langle \gamma^2 \rangle\approx 4\times10^{-4}$. The maximum tolerable LSST multiplicative shear error  of $m<0.003$ produces an error in the integrated power spectrum of $2m \langle \gamma^2 \rangle \approx 2.4\times10^{-6}$.  Spurious shear should not generate power larger than this value, $\langle c^2 \rangle < 2.4\times10^{-6}$, suggesting that we need
\begin{equation}
c_{\rm RMS} < 0.0015.
\end{equation}
If we again allocate half of this error budget to atmospheric dispersion, and keep in mind that the dispersion generates $c_1 = m/2$, we find that the limits on \dv\ induced by additive shear are very similar to those induced by multiplicative shear (Equation (\ref{dvreq})).

One issue that is more severe for additive shear than for multiplicative is that, for multiplicative errors, we are only sensitive to the mean $m$ over all galaxies in a given redshift slice \citep{GuzikBernstein2005}, whereas for $c$ it is the variance across the sky, not the mean value, that is of particular concern.  So we should be alert to aspects of the atmospheric dispersion systematic that will cause spurious shear fluctuating on the $\sim 10\arcmin$ scales that carry most cosmic-shear information.  For instance, if \dv\ depends on source galaxy type within a redshift bin, we must be concerned that there will be variations in $c$ across the field as the mean background type fluctuates.

\subsection{Requirements on atmospheric dispersion errors}
Our final step in setting requirements on \dv\ is to estimate the galaxy size to be used in Equation~(\ref{dvreq}).  The DES sets a goal of 10 lensing source galaxies per arcmin$^2$. 
 {\color{black} \citet{Jouvel} produce a model of the background galaxy population sizes and colors using HST/COSMOS and other observations; from this catalog we note that there are 10~galaxies~arcmin$^{-2}$ with  half-light radii $r\ge 0\farcs4$, hence the {\em largest} possible value of
$\langle 1/r^2\rangle$ for DES would be to use this population.  LSST is more ambitious, with a target source density of 30~arcmin$^{-2}$, which will require reaching down to galaxies with $r\ge 0\farcs27$.  For this latter population, $\langle 1/r^2\rangle \approx (0\farcs35)^{-2}$.  It is also true, however, that many of the galaxies above these cutoff $r$ values will be unusable for weak lensing--- too faint, crowded, or poor photo-z's---which will force the surveys to use galaxies smaller than these best-case cutoff sizes.  As a rough guess, we therefore adopt estimates of $\langle 1/r^2 \rangle = (0\farcs4)^{-2}$ for DES and $(0\farcs27)^{-2}$ for LSST.}
The estimated requirements on \dv\ for the target galaxy population are therefore
\begin{equation}
\label{dvrequirement}
\left|  \left\langle \dv \right\rangle \right| < \left\{
\begin{array}{cc}
6\times10^{-4}\,{\rm arcsec}^2 & {\rm (DES)} \\
1\times10^{-4}\,{\rm arcsec}^2 & {\rm (LSST)}
\end{array} \right.
\end{equation}

Finally we consider requirements on $\dr$, for galaxies where different exposures must be registered for shape measurement.  In this case the mis-registration will cause a blurring of the galaxy that will be roughly equivalent to an added dispersion $\dv \approx \left\langle(\dr)^2\right\rangle$.  The details will depend upon the distribution of filters and zenith angles that one is trying to stack, but we can expect this variation to be of the same order as \dr\ itself.  So we set rough requirements that $\left\langle (\dr)^2\right\rangle$ be below the \dv\ requirements in equation~(\ref{dvrequirement}), or:
\begin{equation}
\label{drrequirement}
\dr_{\rm RMS} < \left\{
\begin{array}{cc}
25\,{\rm mas} & {\rm (DES)} \\
10\,{\rm mas} & {\rm (LSST)}
\end{array} \right.
\end{equation}
An additional point to make here is that two galaxies with opposite signs of \dr\ are both elongated in the same direction after stacking, hence our requirement is not on the population mean of $\dr$; it is on the  RMS value of \dr\ of individual galaxies.  Galaxies with negative \dr\ do not cancel the spurious shear of those with $\dr>0$.

To summarize, atmospheric refraction has the potential to significantly bias the results of cosmic-shear measurements whenever the stellar calibration results in errors for galaxies exceed {\em any} of the following limits:
\begin{enumerate}
\item The dispersion error $\langle \dv \rangle$ of the galaxy population varies with redshift or other binning criteria by an amount exceeding $6(1)\times10^{-4}$~arcsec$^2$ for DES (LSST).  This produces a multiplicative shear error that is too large.
\item The galaxy-to-galaxy variation in \dv\ within a redshift bin substantially exceeds the above values, which could produce an additive shear fluctuating on arcminute scales that biases the cosmic shear power spectrum.
\item The galaxy-by-galaxy error in refraction centroid $\dr_{\rm RMS}$ acquires typical value exceeding 25 (10)~mas in any redshift bin for DES (LSST).  If this occurs, then stacks of images taken at different airmasses and/or filters will be misregistered and both multiplicative and additive shear errors will be produced.
\end{enumerate}

%\newpage

\section{Atmospheric dispersion calculation methods}
\label{calculations}

\begin{figure}%[h!]
\begin{center}
\includegraphics[scale=0.9]{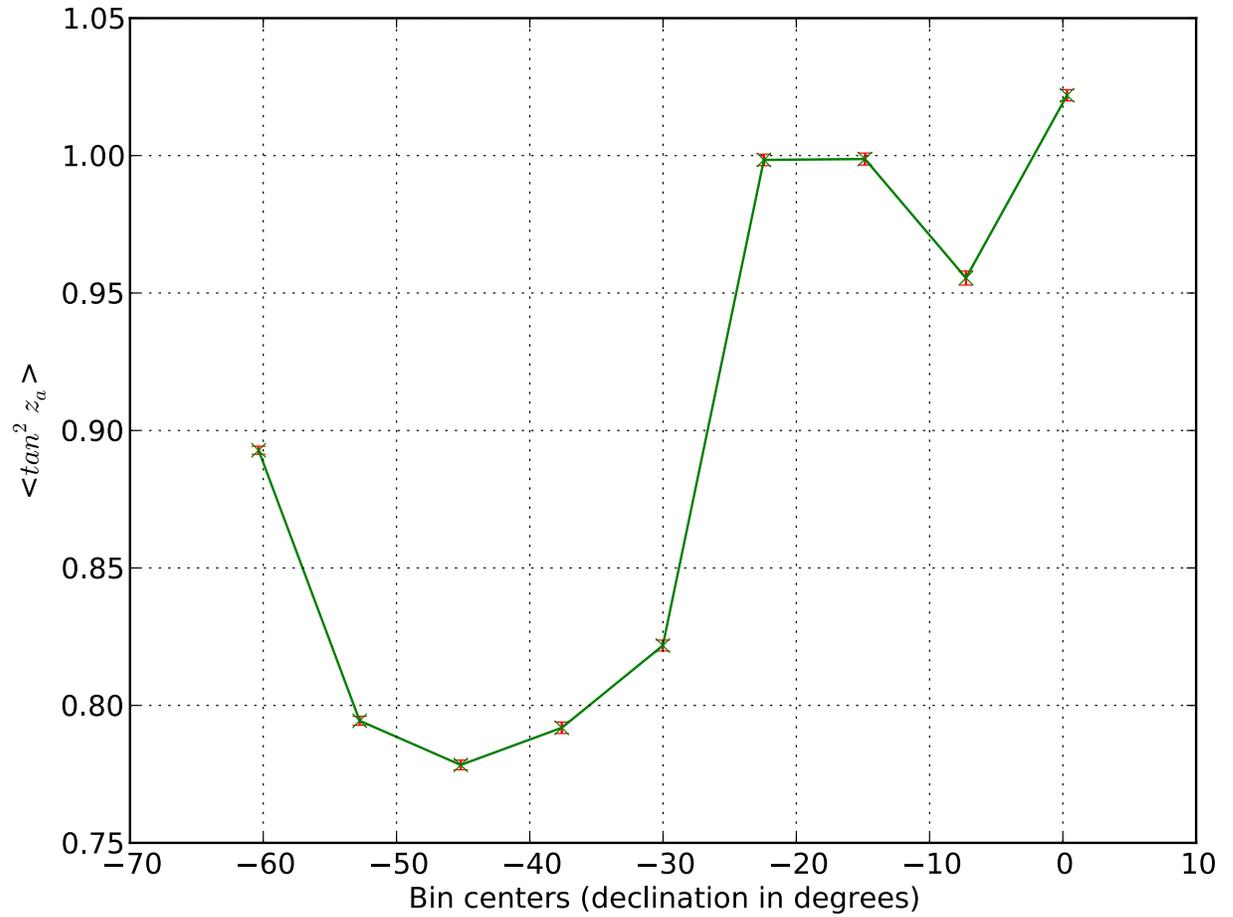}
\caption{ $\tan ^2 z_a$ for {\em DES} exposures binned by field declination. The exposure lists are produced by simulations of the full {\em DES} survey using historical weather patterns and the { \sc ObsTac} survey planning software \citep{neilsen}.}
\label{plotone}
\end{center}
\end{figure}

To quantify the effects of refraction on astronomical images, we first write the refraction angle as 
\begin{equation}
R(\lambda, z_a) = h(\lambda) g(z_a) 
\end{equation}
where $h(\lambda)$ and $g(z_a)$ encode its dependence on wavelength and altitude angle.  As indicated by \citet{cox}, $g(z_a)$ can be approximated as $g(z_a)=\tan (z_a)$.  This functional form implies that \dr\ and \dv\ scale as $\tan (z_a)$ and $\tan^2(z_a) $, respectively. We will present results assuming $z_a = 45\arcdeg$, since they are simple ($\tan 45\arcdeg = 1$) and the \dv\ for a particular field of some survey should be scaled by $\langle \tan^2 z_a \rangle$.  The required scaling factor for DES is not, however, far from unity, so we will ignore it:
we take a simulation of all the DES pointings produced by its scheduling software {\sc ObsTac}\footnote{Provided by Eric Neilsen \citep{neilsen}} and bin them by declination.
Figure~\ref{plotone} shows the relevant quantity $\langle \tan^2 z_a \rangle$ vs declination expected for DES, showing that $\langle \tan^2 z_a \rangle$ is within about 20\% of unity across the declination range of the survey.  For LSST, Figure~3.3 of the 
{\it LSST Science Book}\footnote{http://http://www.lsst.org/lsst/scibook} suggests that the LSST $i$-band observations have $\langle \tan^2 z_a \rangle\approx0.6$, so we can expect that the northern and southern extremes of the survey will have $\langle \tan^2 z_a \rangle$ at or above 0.8, slightly below our nominal value of unity. 

The calculation of shear artifacts induced by \dr\ in a stacking analysis is more complex, because \dr\ is actually a vector quantity and the direction to zenith will vary with hour angle of observations of a given field.  The deleterious effects will scale with the two-dimensional variance of the \dr\ vectors over the course of the survey, so the relevant scaling factor will depend upon the hour-angle distribution as well as zenith-angle distribution, but will be $\le \langle \tan^2 z_a\rangle$.  Our assumption of unity for this scaling factor is hence conservative, but we will see below that the shear systematics from \dv\ are larger than those from \dr\ even with this conservative upper bound on \dr.

 We make use of values for $h (\lambda)$ tabulated in \citet{cox}, slightly corrected to take into account the average conditions of pressure and temperature ($770$ millibar and $11$ C, respectively \citep{stefanik}) at Cerro Tololo, Chile (DES).  As can be readily seen in Equations (\ref{r}) and (\ref{v}), $\overline{R}$ and $V$ both depend on the spectrum of the source object, $S(\lambda)$.  To make our calculations, we use the empirical galaxy Spectral Energy Distribution (SED) templates of  \citet{cww} (CWW) and \citet{kin}\footnote{These spectra are extended in the UV and IR regions by means of the synthetic \citet{bruzual} spectra.}.  The CCW templates consist of 4 galaxies of type E, Im, Scb, Scd and the Kinney templates are representative of both quiescent and starbust galaxies.  For the star spectra, we utilize the stellar libraries provided by \citet{pickles}. We calculate the values of $\overline{R}$ and $V$  for each stellar spectral type,  and for galaxy types at values of redshift $z  \in [0,1]$ with $\Delta z = 0.02$, according to Equations (\ref{r}) and (\ref{v}).  Then, taking a G5V stellar template as our reference mean stellar spectrum, we calculate \dr\ and \dv\ for each object in each of the four $griz$ filters and compare our results to the requirements for cosmic shear derived in \S\ref{effects}.

\section{Results for stellar and galactic spectra}
\label{results}

\begin{figure}%[h!]
\begin{center}
\includegraphics[scale=0.9]{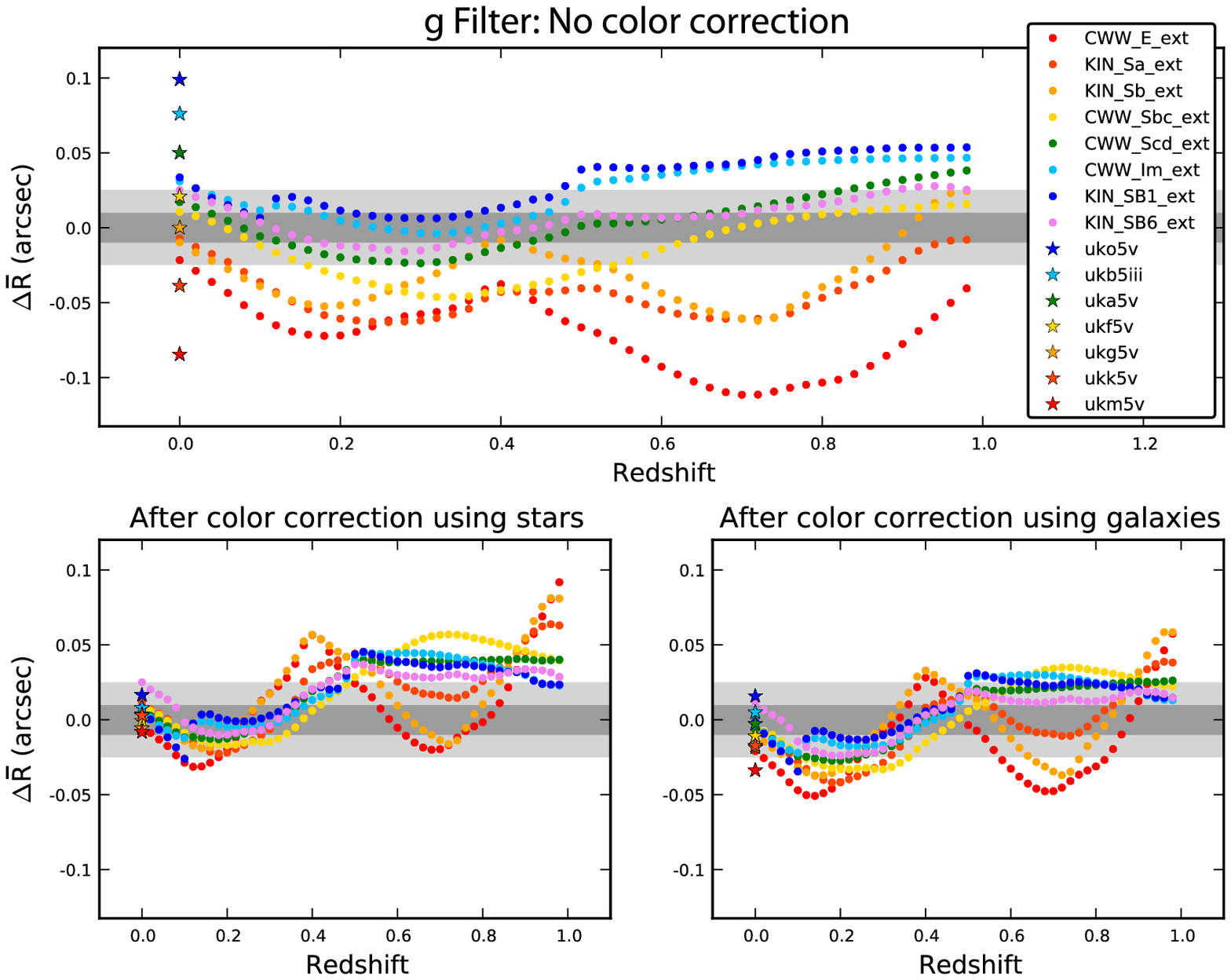}
\caption{ The differential centroid shift \dr\ as a function of redshift ($\Delta z = 0.02$) in the $g$ filter, for zenith angle 45\arcdeg.  Standard templates of galaxies from CWW \citep{cww} (E, Sb, Sc, Im  galaxies ) and Kinney \citep{kin} (quiescent---Sa, Sb---and starburst---SB, SB6---galaxies) were used. \citet{pickles} provides templates of stellar spectra (``ukxxyy" labels). A G5V spectrum was used as the reference to set $\dr=0$. The top panel shows the \dr\ vaues for each spectrum at different values of $z$, whereas the lower panels show the same values after applying a linear correction in color, as described in the text.  DES (LSST) requirements are that $\dr_{\rm RMS}$ fall within the light grey (dark grey) shaded region (Equation (\ref{drrequirement})).}
\label{plottwo}
\end{center}
\end{figure}

\begin{figure}%[h!]
\begin{center}
\includegraphics[scale=0.9]{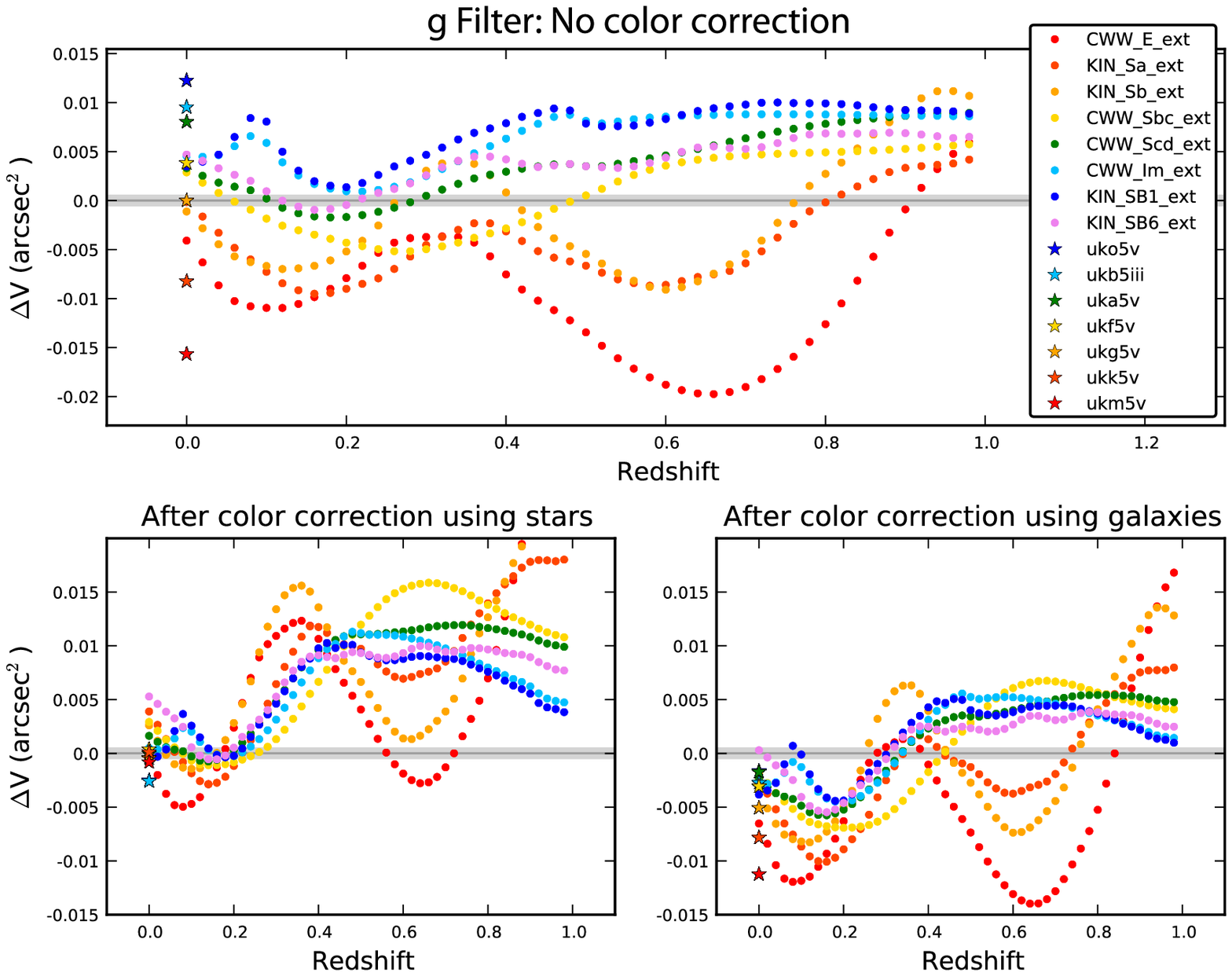}
\caption{ The differential PSF second moment \dv\ as a function of redshift ($\Delta z = 0.02$) in the $g$ filter at zenith angle 45\arcdeg.  Standard templates of galaxies from CWW \citep{cww} (E, Sb, Sc, Im  galaxies ) and Kinney \citep{kin} (quiescent---Sa, Sb---and starburst---SB, SB6---galaxies) were used. \citet{pickles} provides templates of stellar spectra (``ukxxyy" labels). A G5V spectrum was used as the reference to define $\dv=0$.  The top panel shows the \dv\ vaues for each spectrum at different values of $z$, whereas the lower panels show the same values after applying a linear correction in color, as described in the text.  DES (LSST) requirements are that \dv\ fall within the light grey (dark grey) shaded region (Equation (\ref{dvrequirement})).} 
\label{plotthree}
\end{center}
\end{figure}

The top panels of Figures \ref{plottwo} and \ref{plotthree} show the calculated values of \dr\ and \dv\ in the $g$ filter as a function of redshift (at $z_a = 45\arcdeg$). The remaining panels show these same quantities after applying a correction linear in color that will be described in the next section. Similar calculations and plots were performed and obtained in the $r$, $i$, and $z$ filters. 
 The results for stellar spectra at $z=0$ show how the effect of atmospheric refraction is greater for bluer sources, as expected. DES (LSST) requirements, as calculated in Equations (\ref{dvrequirement}) and (\ref{drrequirement}), are also shown in the plots. To determine whether these requirements are satisfied, we first examine the galaxies expected to be detected by the DES according to the DES Data Challenge 6B.\footnote{Simulations created by the DES Data Management Project http://cosmology.illinois.edu/DES/main/}  To each of the $\approx$250,000 galaxies detected in a 10~deg$^2$ subsample of the simulation we assign a \dv\ and \dr\ from the template that most closely matches its colors and redshift.  Then we divide the galaxy population into four redshift bins, computing the mean value and the dispersion.
of \dv\ and the RMS  value of $\dr$ among the galaxy population.  These are represented by the blue circular dots in Figures \ref{plotsix} and \ref{plotseven} (we assume that the LSST population has similar type distribution to the DES population). 

It can be seen that cosmic shear measurements performed in $g$ band will, in the absence of some correction scheme, have systematic errors due to atmospheric dispersion that dominate the statistical errors, with \dv\ up to $6\times$ our threshold of importance for DES and $30\times$ for LSST. The $\dr_{\rm rms}$ values are also large enough to cause dominant systematic errors, but \dv\ induces worse problems, so we will focus on \dv\ henceforth.  In $r$ band, the levels of \dv\ and $\dr_{\rm RMS}$ remain large enough to dominate the LSST error budget but are $1.6\times$ our nominal threshold for DES systematic errors, so may be acceptable for DES.  For LSST, even in $i$ band, the dispersion effects are up to $2.6\times$ larger than our derived requirements, so might be subdominant to statistical errors if our various approximations have been too conservative.  But certainly the refraction effect requires attention in constructing an LSST error budget.  Atmospheric dispersion is not a problem for either experiment in $z$-band (and we have verified that this is also true, as expected, in $Y$ band).

\begin{figure}%[h!]
\begin{center}
\includegraphics[scale=0.9]{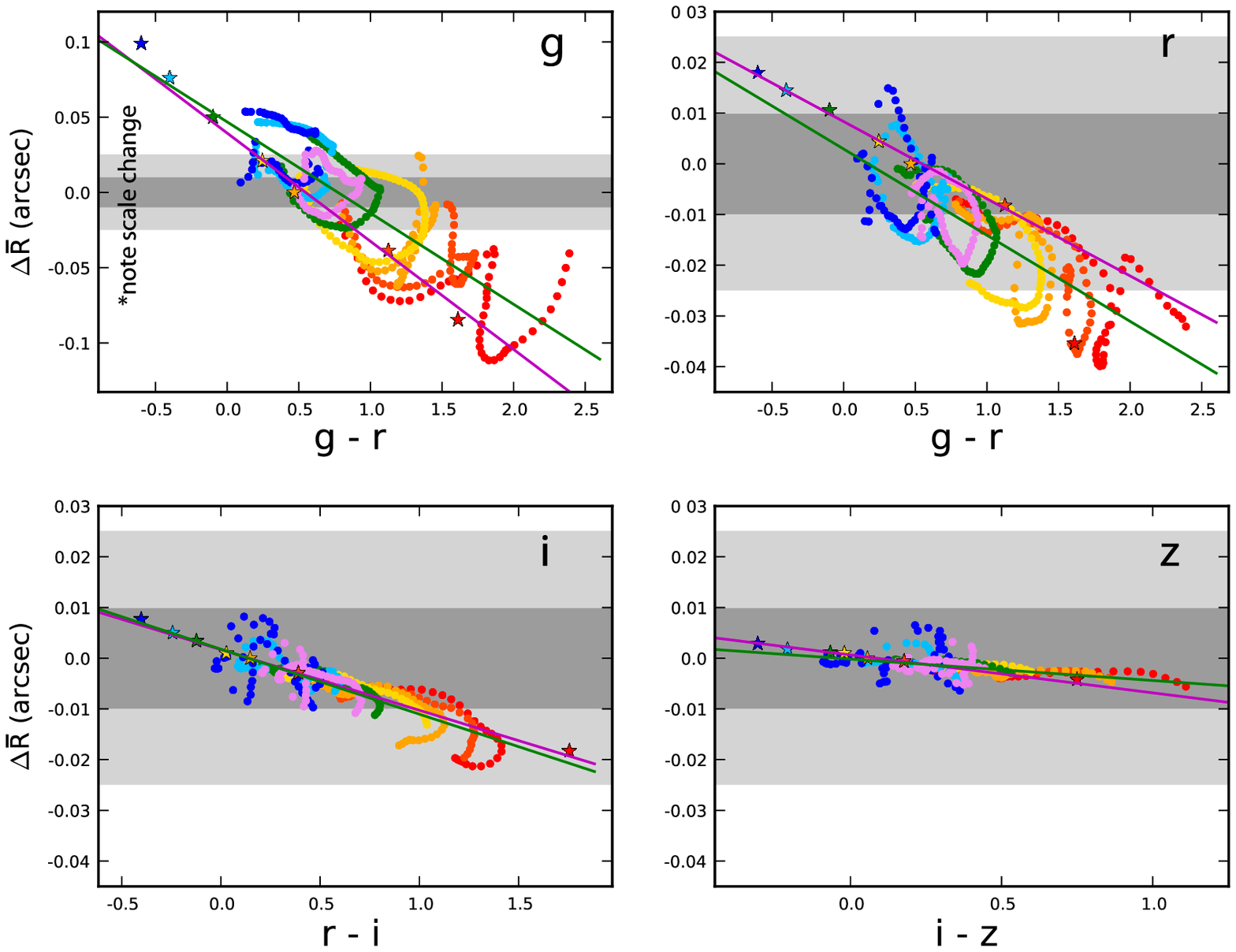}
\caption{\dr\ as a function of color for stellar and galaxy spectra at different redshift values ($\Delta z= 0.02$) for each of the $griz$ filters at $z_a=45\arcdeg$.  The solid green (magenta) line represents a linear fit using only galaxy (stellar) spectra. Requirements for DES (LSST) on $\dr_{\rm RMS}$ are shown by the light grey (dark grey) shaded regions.}
\label{plotfour}
\end{center}
\end{figure}

\begin{figure}%[h!]
\begin{center}
\includegraphics[scale=0.9]{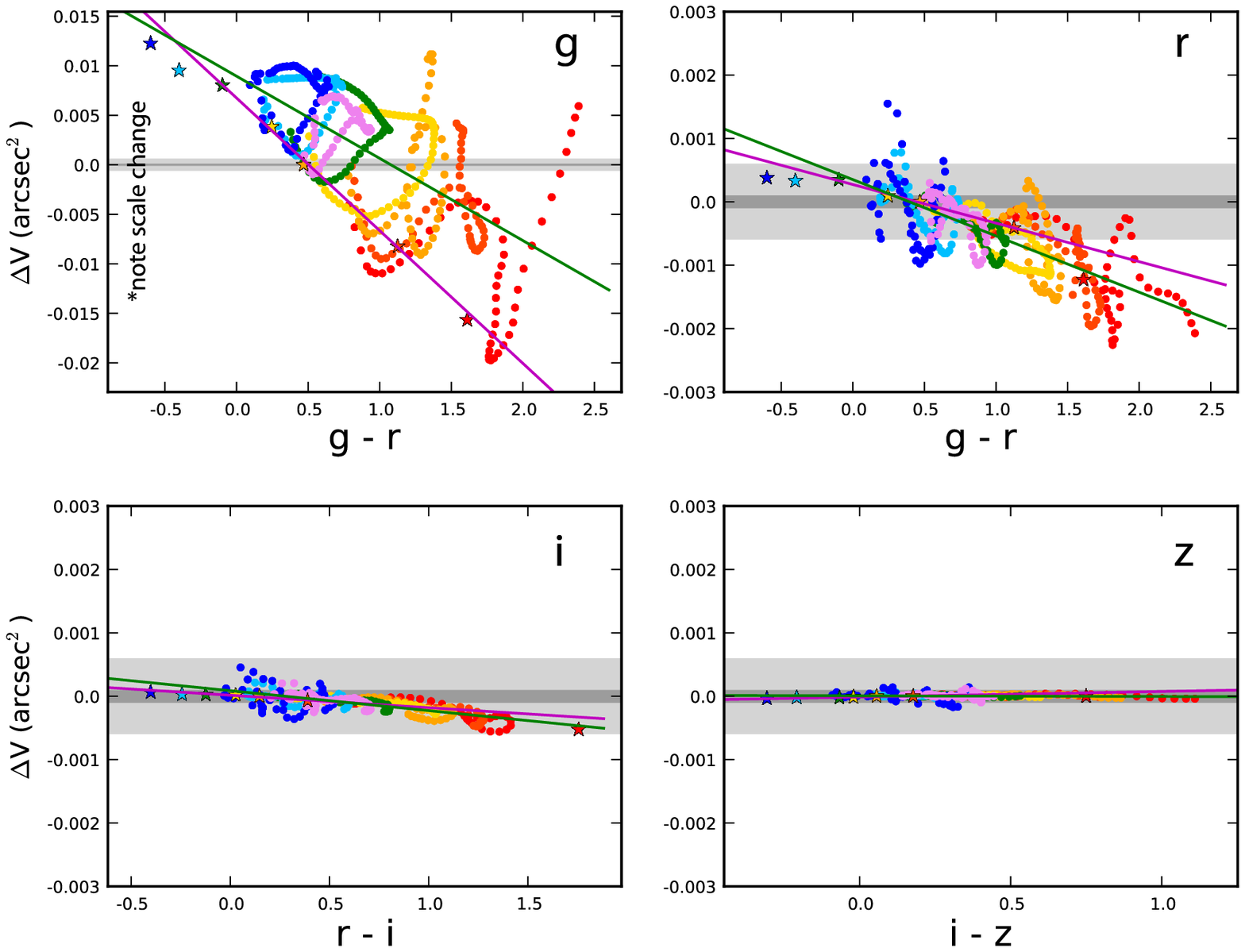}
\caption{\dv\ as a function of color for stellar and galaxy spectra at different redshift values ($\Delta z= 0.02$) for each of the $griz$ filters at $z_a=45\arcdeg$. The solid green (magenta) line represents a linear fit using only galaxy (stellar) spectra. Requirements for DES (LSST) on \dv\ are shown by the light grey (dark grey) shaded regions.}
\label{plotfive}
\end{center}
\end{figure}

\section{ Strategies for calibrating effects of atmospheric dispersion}
\label{corrections}
\begin{figure}%[h!]
\begin{center}
\includegraphics[scale=0.9]{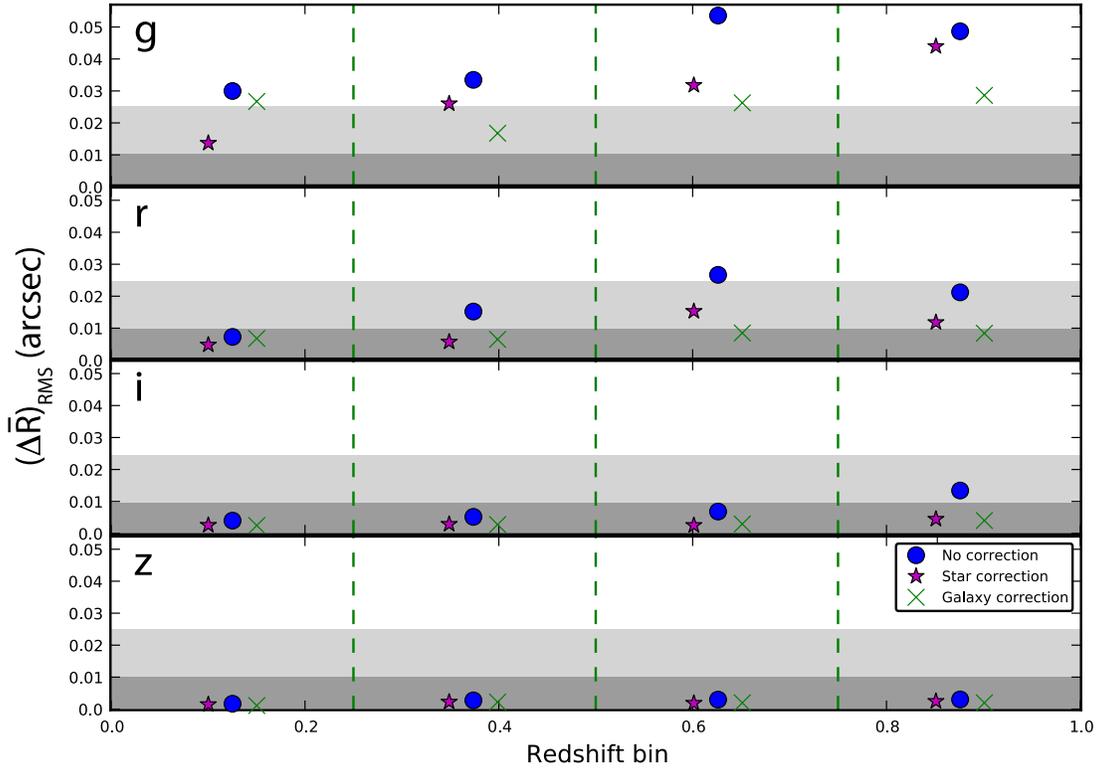}
\caption{$\dr_{\rm RMS}$ in four equally-spaced redshift bins for each of the four $griz$ filters when no correction (blue dots), a  linear correction in color using only star spectra (magenta stars) and a linear correction using only galaxy spectra (green crosses) is applied. DES (LSST) requirements are that $\dr_{\rm RMS}$ fall within the light grey (dark grey) shaded region (Equation (\ref{drrequirement})).} 
\label{plotsix}
\end{center}
\end{figure}

\begin{figure}%[h!]
\begin{center}
\includegraphics[scale=0.9]{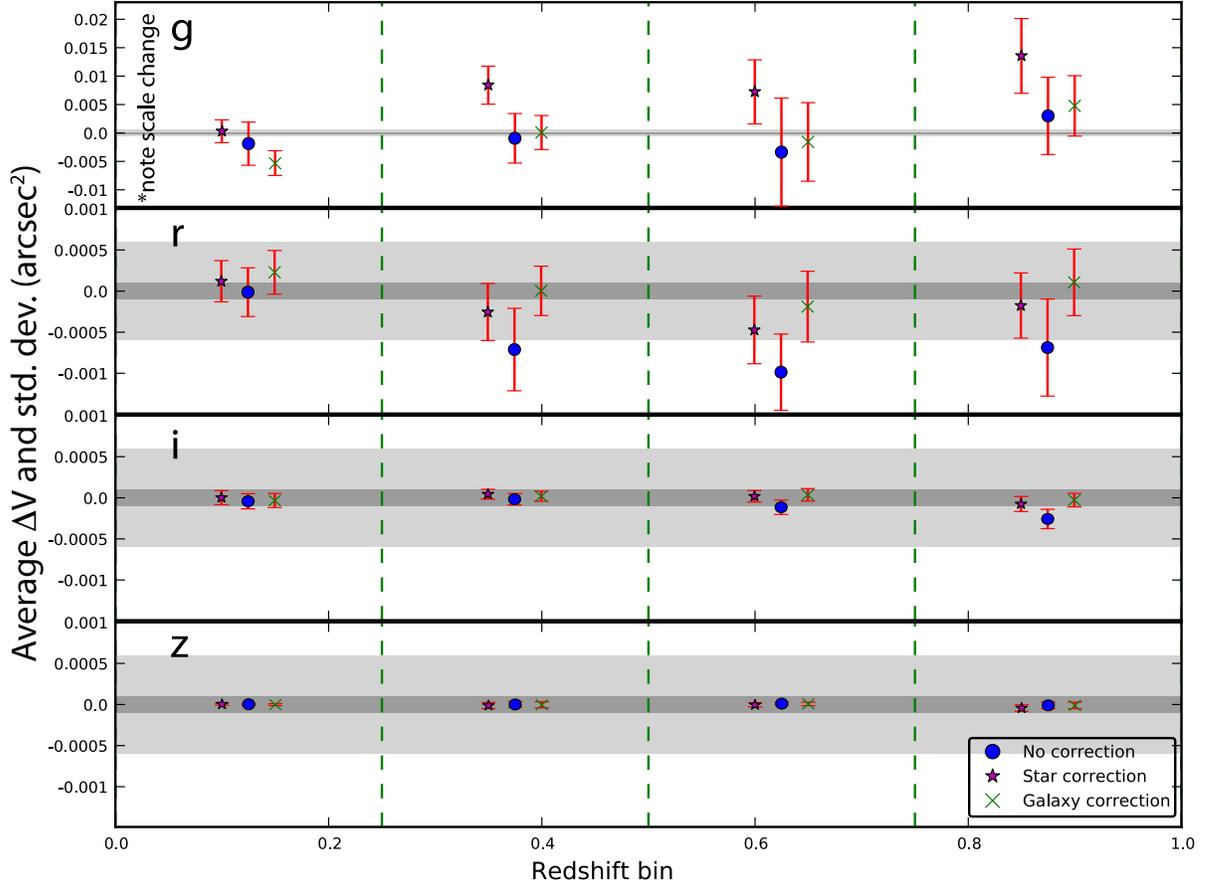}
\caption{ \dv\ in four equally-spaced redshift bins for each of the four $griz$ filters when no correction (blue dots), a  linear correction in color using only star spectra (magenta stars), and a linear correction using only galaxy spectra (green crosses) is applied.  The vertical bars depict the standard deviation of \dv\ between different galaxy templates and redshifts within each bin.  They therefore represent the range of variation that the mean \dv\ might accrue due to variations in the population mix of galaxies with redshift and with position on the sky.   
DES (LSST) requirements are that \dv\ fall within the light grey (dark grey) shaded regions (Equation (\ref{dvrequirement})).} 
\label{plotseven}
\end{center}
\end{figure}

The previous section shows that the effects of atmospheric dispersion on shape measurements are non-negligible in the $g$, $r$ and $i$ filters. As was pointed out above,  \dr\ and \dv\ depend on the spectra of the observed objects, which cannot be measured in a photometric survey like DES.  We would like to develop a calibration or correction that depends only on observable quantities, such as the color of the stars and galaxies. We calculate \dr\ and \dv\ of each star and galaxy at different redshift values and filters, and plot them as a function of color in Figures \ref{plotfour} and \ref{plotfive}. The plots suggest that a function of color can be defined that partially corrects the values of  \dr\ and \dv.  We determine \dr\ and \dv\ functions that depend linearly on color and minimize the RMS residuals of the stellar templates---this is an operation that could be done empirically from the observations. The stars in Figures \ref{plotsix} and \ref{plotseven} give the $\dr_{\rm RMS}$, and mean and variance of \dv, for the galaxy templates in 4 redshift bins after application of the star-based linear color correction to the galaxies. A linear function of color that minimizes the RMS residuals of \dr\ and \dv\ for the {\em galaxy} templates (crosses in Figures \ref{plotfive} and \ref{plotsix}) produces a calibration with smaller residuals for galaxies, but note that such a calibration would have to rely on theoretical calculations since \dv\ cannot be measured from the data for the galaxies. The linear fits to both stars and galaxies are also shown in Figures \ref{plotfour} and \ref{plotfive}. The $\dr_{\rm RMS}$ and \dv\ for the galaxies after applying the linear corrections at each filter are shown in Figures \ref{plotsix} and \ref{plotseven}. Either  linear correction brings the atmospheric dispersion effects within requirements in the $r$ ($i$) band for DES (LSST).  The LSST $r$ band systematic errors are still above our thresholds of concern after linear color corrections, by factors up to 5 depending on the redshift bin and type of correction---thus must still be considered an important contributor to the LSST error budget, although suggestive that more sophisticated correction schemes could bring atmopheric-dispersion systematics back below statistical errors in this case. For $g$ band, however, the linear corrections still leave errors well above our thresholds in both DES and LSST, and in fact the ``corrected'' data are worse in most of the redshift bins than uncorrected ones. Also we note that the galaxy-to-galaxy standard deviation of \dv\ within a redshift bin is typically much larger than the DES or LSST requirements we have derived, suggesting that spatial variations of the induced additive shear will be an issue. It appears that atmospheric dispersion is likely to create dominant systematic errors in $g$ band, and the surveys should plan on basing cosmic-shear measurements on the redder bands. We have verified that $u$ band is, as expected, even worse than $g$.

\newpage
\section{Conclusion}
\label{conclusion}
Atmospheric dispersion can have a significant effect in PSF correction of shape measurements in broadband photometric surveys by affecting the first and second moments of the observed images.  If uncorrected, it produces cosmic-shear spurious signals larger than the desired statistical uncertainties for the $g$ and $r$ bands of the DES, and for the $g$, $r$, and $i$ bands for LSST.  This effect depends on the spectrum of the source and therefore is not easy to correct.  A simple linear correction to the PSF position and size based on the observed color, calibrated empirically from stars or estimated theoretically for galaxies, should improve calibration of dispersion effects significantly.  Even after applying this correction, however, weak lensing measurements in the $g$ band are significantly degraded by dispersion effects for both DES and LSST.
Cosmic shear measurements can be performed with images taken in the $r$ band for the accuracy required by DES, but further work will be needed in the case of LSST.  Shape and shear measurements in the $z$ and $Y$ band can safely be performed without attention to dispersion corrections.

There are several possible tactics for further reduction of cosmic-shear systematic errors from atmospheric dispersion:
\begin{itemize}
\item Use an atmospheric dispersion corrector (ADC) during the imaging.  While this option is available for KIDS cosmic-shear surveys being conducted on the VLT Survey Telescope\footnote{http://www.eso.org/public/teles-instr/surveytelescopes/vst.html} and for the HyperSuprime Camera to be installed on Subaru\footnote{http://www.naoj.org/}, an ADC is not part of the optical designs for DES and LSST.  
\item Observe at  lower zenith angles---\dr\ and \dv\ scale as  $\tan (z_a)$ and $\tan^2(z_a)$, respectively, so would be reduced by factors of $\sqrt 3$ and 3, respectively, if we were to take nominal zenith angles of 30\arcdeg\ instead of the 45\arcdeg\ assumed in our calculations.  In practice it would not be easy to maintain such low airmasses throughout a large survey, especially if one wants to survey 15,000~deg$^2$ or more of low-Galactic-latitude sky from the latitudes of $\approx -30\arcdeg$ of both Cerro Tololo (DES) and Cerro Pach\'on (LSST).   
\item Derive a more elaborate correction than the linear fit we have used, {\it e.g.}, an approach that makes use of multi-color photometry to estimate the full spectrum of each galaxy---which of course is already done in the course of most template-fitting estimators for photometric redshifts.  This technique is discussed by \citet{Cypriano} in the context of calibrating wavelength-dependent PSF variation across the very broad band planned for the Euclid spacecraft.  In the Euclid case it is telescope diffraction, not atmospheric refraction, that causes the wavelength dependence of the PSF, but the issue is similar. 
\item One could simulate the full effect in detail to estimate additive and multiplicative corrections that must be applied to the measured shear fields. This approach would have the usual limitations of precision calibration by simulation: it would be accurate only to the extent that the simulation captures the full multivariate distribution of the sizes, shapes, redshifts, and SEDs of galaxies in the true sky. The SED distribution will be particularly challenging to constrain for the faint end of the surveys' target galaxy populations.
\end{itemize}
It seems likely, therefore, that both DES and LSST can make use of $r$ or redder bands for cosmic-shear surveys, but this will require explicit galaxy-by-galaxy correction for PSF and astrometric dependence on the source spectrum.  For DES, we show that a simple linear color correction will suffice, but more work is needed to devise a correction that recovers the full cosmic-shear utility of the LSST $r$ band data.

We note lateral color in telescope optics causes spurious spectrum-dependent cosmic-shear signals in a manner very similar to the atmospheric dispersion that we examine here.  As with atmospheric dispersion, it is typical to specify an optical design that keeps this contribution well below the size of the PSF, but this is not a sufficient condition to guarantee that differences between galaxy and stellar spectra produce PSF calibration errors below the requirements of cosmic-shear surveys.  

\acknowledgments
This work was supported by NSF grant AST-090827 and DOE grant DE-FG02-95ER40893.

%\newpage

%\bibliographystyle{unsrtnat}
\bibliographystyle{plainnat}

\bibliography{ms}

\end{document}